\documentclass[prd,twocolumn,floatfix,superscriptaddress,nofootinbib]{revtex4-1}
\usepackage{dcolumn,amsmath}
\usepackage{graphicx}
\usepackage{braket}
\usepackage{bm}
\usepackage{amssymb}
\usepackage{hyperref}
\usepackage{multirow}
\usepackage{footnote}
\usepackage{subcaption}
\usepackage{ragged2e}
\usepackage{xcolor}
\usepackage{pdfcomment}
\usepackage[justification=justified,labelfont=bf,singlelinecheck=off]{caption}
\usepackage{threeparttable,booktabs}  
\usepackage{subcaption,siunitx,booktabs}
\usepackage{caption,afterpage,tabularx}
\usepackage{diagbox}
\usepackage{mathrsfs}
\usepackage{physics}
\usepackage{cancel}
\usepackage{amsmath}
\usepackage{amssymb}
\usepackage{color}
\usepackage{CJKutf8}

\newcommand{\JSU}{
School of Physics and Electronic Engineering,
\\Jiangsu University, Zhenjiang, 212013 Jiangsu, China\\
}

\begin{document}

%\begin{CJK}{UTF8}{gbsn}%
\title{Dark photons from dineutron decays in neutron stars}

\author{Yongliang Hao}
\email{yhao@ujs.edu.cn}
\affiliation{\JSU}
\author{Zhenwei Chen}
\affiliation{\JSU}

\date{\today}

\begin{abstract}

We focus on a novel baryon-number ($\mathcal{B}$) violating process within neutron stars, where two neutrons convert into two dark photons ($nn \rightarrow VV$) via new Higgs-like scalar bosons. This process is believed to be greatly suppressed at low energies but could be highly amplified in a dense neutron environment like neutron stars. The $nn \rightarrow VV$ process could give rise to non-trivial effects that are distinct from similar processes in previous studies and could alter the properties of neutron stars, such as orbital periods, collapse thresholds, stability conditions, cooling rates, gravitational wave emissions, etc. The emitted dark photons may serve as dark-matter candidates and exhibit special red-shifted energy spectra mainly linked to the compactness of the neutron star. We point out that the dark photons emitted from neutron stars may yield detectable signals in future experiments. We also show that the precision pulsar‑timing data provides a powerful tool to constrain the parameter space of new-physics models. The study of the $nn \rightarrow VV$ process, which combines astronomical observations and particle physics models together, may open new windows into the detection of the $\mathcal{B}$-violating effects and may also provide new insights on the study of dark matter.

\end{abstract}

\maketitle

\section{Introduction}
The Standard Model (SM) of particle physics is considered as a successful theory in describing the fundamental particles and their interactions excluding gravity \cite{navas2024review}. An important achievement of the SM is the discovery of the SM-like Higgs boson \cite{aad2012observation,chatrchyan2012observation,chatrchyan2013observation}. In spite of its success, there remain many pending questions that cannot be explained in a satisfactory way by the SM. Among them, the matter-antimatter asymmetry, which is characterized by the observed excess of matter over antimatter in the universe, is still one of the main questions \cite{navas2024review}.

Baryon number violation (BNV) is one of the three criteria suggested by Sakharov to account for the observed asymmetry between matter and antimatter\cite{sakharov1967violation}. $\mathcal{B}$-violation is anticipated to exist in a wide variety of modes. For example, the dineutron decay with invisible final states ($nn \rightarrow \text{inv.}$) has been theoretically predicted by numerous new-physics models \cite{heeck2020inclusive,girmohanta2020improved,girmohanta2020baryon,he2021eft} and extensively investigated in various experimental studies \cite{bernabei2000search,araki2006search,anderson2019search,allega2022improved}. 
From an experimental perspective, several experiments have reported the limits on the partial lifetimes associated with dineutron decay modes that have invisible final states ($T_{nn \rightarrow \text{inv.}}$), such as SNO+ ($1.3 \times 10^{28}$ yr \cite{anderson2019search}, $1.5 \times 10^{28}$ yr \cite{allega2022improved}), KamLAND ($1.4 \times 10^{30}$ yr \cite{araki2006search}), LNGS ($1.2 \times 10^{25}$ yr \cite{bernabei2000search}), etc. Furthermore, the JUNO experiment has a potential for an improved sensitivity in the future ($1.4 \times 10^{32}$ yr \cite{juno2025juno}).

Dark photons ($V$) are hypothetical bosons that may arise in the new-physics models with extra $U(1)_D$ gauge symmetries \cite{navas2024review}. They barely interact with the SM particles and could potentially be viable candidates for dark matter \cite{fabbrichesi2021physics,barducci2021dark,Hosseini2022collider,Abdullahi2023Semivisible,Capanelli2024Gravitational}. Direct and indirect searches for dark photons have been ongoing for many years \cite{araki2021dark,cheung2024probing,graham2021searches,Adrian2023JLab,Romanenko2023SRF,Abreu2024FASER,Gil2024NA62,Knirck2024Antenna,Tang2024Cavity,linden2024indirect,linden2024xray}. Given that dark photons have minimal interaction with ordinary matter, it remains very difficult to detect them with currently available experimental techniques (see e.g. Refs. \cite{Knirck2024Antenna,Tang2024Cavity}). The dineutron decay into two dark photons ($n n \rightarrow V V$) violates two units of $\mathcal{B}$ ($|\Delta \mathcal{B}| = 2$) and possesses numerous intriguing signatures that distinguish it from other decay modes with invisible final states. The $n n \rightarrow V V$ process is characterized by the disappearance of two neutrons and the appearance of two dark photons. Since the new-physics energy scale tends to significantly suppress the decay rate for the $nn \rightarrow VV$ process, the detection of dark photons from this process in a terrestrial laboratory faces unprecedented challenges.

Neutron stars, among the densest objects in the universe, are considered as neutron-rich environments, where neutron-related BNV processes might occur at a more significant rate \cite{berryman2022neutron}. The specific BNV process, i.e. $nn \rightarrow VV$, can be mediated by additional Higgs-like scalar bosons through high-dimensional effective operators at the quark level. This process is believed to be greatly suppressed at low energies and has not been confirmed in the currently ongoing experiments (see, e.g. Ref. \cite{pascual2020lhc}). In contrast, the number of neutrons contained in a neutron star could be so large that the new-physics effects associated with the $nn \rightarrow VV$ process can be greatly enhanced, offering a good opportunity to explore new physics beyond the SM.

This work shares similar background motivation, methodological structure, and astrophysical assumptions with a previously study by the same first-author in Ref. \cite{hao2023dineutron}. However, the present work focuses on a distinct dineutron decay channel involving dark photons and in particular includes new physical signatures, distinct particle phenomenology, and updated parameter constraints that are different from the previous study. Again, we first examine the new physics models that may give rise to the $nn \rightarrow VV$ process mediated by new scalar bosons. The dineutron decay rate for the $nn \rightarrow VV$ process is then estimated using these models. The structure of neutron stars and the equation of state (EoS) for neutron-star matter are then briefly reviewed. Following that, we focus on the observable effects of the $nn \rightarrow VV$ process on neutron star characteristics, including orbital-period anomaly and dark-photon emission. We shall use the natural units (i.e., $c \equiv 1$, $\hbar \equiv 1$) in this article unless otherwise noted.

\section{The model}

\begin{figure}[b] 
\centering
\includegraphics[scale=0.99, width=0.99\linewidth]{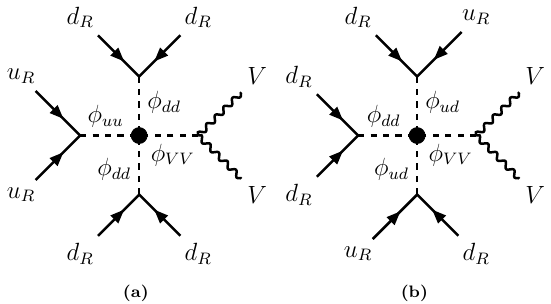}
\caption{Possible tree-level diagrams for the $n n\rightarrow V V$ mediated by the additional scalar bosons at the quark level.
}
\label{fig01}
\end{figure}

Figure \ref{fig01} shows possible diagrams at the tree level for the dineutron decay into dark photons ($n n\rightarrow V V$) mediated by the new scalar bosons, such as diquarks \cite{mohapatra1982hydrogen,mohapatra1983spontaneous}, etc. The new scalar bosons can be accommodated in partially unified models with the symmetry group $SU(4)_c \times SU(2)_L \times SU(2)_R$, such as the Pati-Salam model \cite{pati1974lepton,pati1975erratum} and its variants \cite{mohapatra1980local,davidson1979b}. Such models are characterized by treating quarks and leptons on the equal footing. The left-right symmetric (LRSM) model based on the symmetry group $SU(3)_c \times SU(2)_L \times SU(2)_R \times U(1)_{B-L}$ \cite{pati1974lepton,mohapatra1975left,mohapatra1975natural,senjanovic1975exact} is a low-energy effective model that can be embedded in the partially unified models. Considering symmetry properties of the partially unified models and less constrained couplings, we assume that the diquarks tend to couple to the right-handed fermions, which transform as a singlet under $SU(2)_L$ and as a doublet under $SU(2)_R$ (see e.g. Refs. \cite{mohapatra1980local,babu2009neutrino,patra2014post}). Under the LRSM symmetry group, the right-handed quarks of the first generation transform as \cite{mohapatra1982hydrogen,mohapatra1983spontaneous},
\begin{equation}
\begin{split}
q_R\Bigl(3, 1, 2,\frac{1}{3}\Bigr) &
 =
 \left(
  \begin{array}{c}
   u\\
    d
   \end{array}
   \right)_R.
 \end{split}
\end{equation}
Here, the subscript $R$ represents the right-handed spinor defined by $q_{R}\equiv (1+ \gamma^5)q/2$. The new scalar bosons that mediate the $n n\rightarrow V V$ process can be written as \cite{mohapatra1982hydrogen,mohapatra1983spontaneous,bolton2019alternative,nieves1984analysis,chen2011type,de2019implementing}
\begin{equation}
\begin{split}
\phi^{(R)}_q\Bigl(\bar{6},1, 3,-\frac{2}{3}\Bigr)
&=
\left(
  \begin{array}{cc}
   \frac{\phi_{ud}}{\sqrt{2}}         &  \phi_{dd}\\
   \phi_{uu} & -\frac{\phi_{ud}}{\sqrt{2}}
  \end{array}
  \right)_R.
\end{split}
\end{equation}

Considering the fact that the scalar boson in the SM (i.e. the Higgs boson) couples to the neutral gauge boson of the weak interactions via the term $M_ZhZ_{\mu}Z^{\mu}$, we assume that the remaining new scalar boson ($\phi_{VV}$) couples to the massive dark photon via the new term $M_V\phi_{VV}V_{\mu}V^{\mu}$. The $nn \rightarrow VV$ process can be achieved by combining this new term with the traditional term mediated by diquarks. Inspired by Refs. \cite{mohapatra1980local,barbieri1981spontaneous,mohapatra1982hydrogen,mohapatra1983spontaneous,babu2009neutrino,babu2012coupling}, we assume that the relevant operators that are responsible for the $nn \rightarrow VV$ process depicted in Fig. \ref{fig01} can be written as 
\begin{equation}
\begin{split}
O_{s}  \equiv & g_{\alpha \beta} q^{\alpha T}_{R} C^{-1} i \sigma_2 \phi_q q^{\beta}_{R} + g_V M_V \phi_{VV} V_{\mu}V^{\mu}\\
+ & f_{\phi} \epsilon_{ikm} \epsilon_{jln} \phi_{dd}^{ij} \phi_{dd}^{kl} \phi_{uu}^{mn} \phi_{V V }\\
+  & \text{H.c.}
\end{split}
\label{fgdefinition}
\end{equation}
or,
\begin{equation}
\begin{split}
O_{s}  \equiv & g_{\alpha \beta} q^{\alpha T}_{R} C^{-1} i \sigma_2 \phi_q q^{\beta}_{R} + g_V M_V \phi_{VV} V_{\mu}V^{\mu}\\
+ & f_{\phi} \epsilon_{ikm} \epsilon_{jln} \phi_{ud}^{ij} \phi_{ud}^{kl} \phi_{dd}^{mn} \phi_{V V }\\
+  & \text{H.c.}
\end{split}
\label{fgdefinition}
\end{equation}
Here, we have extracted the mass of the dark photon ($M_{V}$) from the new term so that $g_V$ is a dimensionless coupling constant. The parameters $g_{\alpha \beta}$, $f_{\alpha \beta}$ and $f_{\phi}$ are also dimensionless coupling constants. The charge conjugation operator is denoted by $C$. The $SU(3)_c$ and $SU(2)_R$ indices are denoted by Latin ($i$, $j$, $k$, $l$, $m$ and $n$) and the Greek letters ($\alpha$ and $\beta$), respectively.

For simplicity, we prefer to present our results at the nucleon level. We assume that the above-mentioned coupling constants associated with the color and flavor of quarks can be absorbed into a few coupling constants (i.e. $g_1$ and $g_2$ defined below). Such an assumption is similar to the way of defining the form factors of nucleons, where the main properties of the quark-level interactions and the general information of the nucleon structure can be encapsulated into several parameters without paying attention to all the details of the interactions. This trick has been widely applied in the description of the interaction between an elementary particle and a composite particle in nuclear physics. Actually, this assumption makes sense because we can always encapsulate the details of quark-level interactions into the coupling parameters by carefully adjusting their values without causing any inconsistencies with the present experimental data.

At the nucleon level, the $nn \rightarrow VV$ process can be effectively described by 
\begin{equation}
\begin{split}
\mathscr{L} \supset& -\frac{1}{4} F^{\prime}_{\mu \nu} F^{\prime \mu \nu} + \frac{1}{2}M_V^2 V_{\mu}V^{\mu} + \frac{1}{2} \partial_{\mu} \phi \partial^{\mu} \phi \\
& - \frac{1}{2}M_{\phi}^2 \phi^2 +\bar{n}(i\cancel\partial-m_n)n + g_2 M_V \phi V_{\mu}V^{\mu} \\
& + g_1 \bar{n}n^c \phi + \text{H.c.}
\end{split}
\end{equation}
Here, $g_1$ and $g_2$ are the coupling constants of the new scalar boson to the neutron and dark photon, respectively. $m_{n}$ is the mass of the neutron. $M_{\phi}$ is the mass of the new scalar bosons and can be interpreted as the energy scale of new physics.

The phenomenological constraints on the coupling constants ($|g_1 g_2|$) and the mass  ($M_{\phi}$) of the new scalar bosons depend on the choice of experimental data and specific theoretical models. Currently, there is no direct experimental information on $M_{\phi}$ and $|g_1 g_2|$. Some useful insights or clues on the strengths of such parameters could be found from the studies of precision atomic or molecular spectroscopy. Anomalous changes in atomic or molecular transitions (e.g. isotope shifts \cite{delaunay2017probing,mikami2017probing,berengut2018probing,brax2018bounding,berengut2020generalized,tanaka2020relativistic,counts2020evidence,rehbehn2021sensitivity,hur2022evidence,figueroa2022precision,rehbehn2023narrow,delaunay2023self,chang2024systematic,wilzewski2024nonlinear,door2025probing}) beyond theoretical predictions can potentially arise from new bosons and have been comprehensively studied (see e.g. Refs. \cite{delaunay2017probing,mikami2017probing,berengut2018probing,brax2018bounding,berengut2020generalized,tanaka2020relativistic,counts2020evidence,rehbehn2021sensitivity,hur2022evidence,figueroa2022precision,rehbehn2023narrow,delaunay2023self,chang2024systematic,wilzewski2024nonlinear,door2025probing,banks2021charting,frugiuele2022muonic,liu2025probing}). The measurements of such transitions provide a useful and supplementary tool for testing the SM and constraining the parameter space for new interactions beyond the SM \cite{delaunay2017probing,mikami2017probing,berengut2018probing,brax2018bounding,berengut2020generalized,tanaka2020relativistic,counts2020evidence,rehbehn2021sensitivity,hur2022evidence,figueroa2022precision,rehbehn2023narrow,delaunay2023self,chang2024systematic,wilzewski2024nonlinear,door2025probing,banks2021charting,frugiuele2022muonic,liu2025probing}. As a distinctive feature of the anomalous atomic (molecular) transitions, the corresponding interactions can be described by a Yukawa-type potential \cite{delaunay2017probing,mikami2017probing,berengut2018probing,berengut2020generalized,tanaka2020relativistic,counts2020evidence,rehbehn2021sensitivity,hur2022evidence,figueroa2022precision,rehbehn2023narrow,delaunay2023self,chang2024systematic,wilzewski2024nonlinear,door2025probing}, which is parameterized by the coupling parameter to nucleons ($y_N$) and to electrons ($y_e$). The derived bounds on the product of such parameters depend on the mass of the new bosons and roughly lie within a very broad range from $10^{-10}$ to $10^{-25}$ (i.e. $|g_N g_e| \lesssim  10^{-10}$-$10^{-25}$) or even a broader range in the literature \cite{delaunay2017probing,mikami2017probing,berengut2018probing,berengut2020generalized,tanaka2020relativistic,counts2020evidence,rehbehn2021sensitivity,hur2022evidence,figueroa2022precision,rehbehn2023narrow,delaunay2023self,chang2024systematic,wilzewski2024nonlinear,door2025probing,banks2021charting,frugiuele2022muonic,liu2025probing}, making it difficult to compare such bounds because they are based on different theoretical models and different experimental configurations. Nevertheless, such bounds could provide useful reference values for the values of $|g_1 g_2|$. Or at least, we could predict the possible values of $|g_1 g_2|$ by appealing analogies with the atomic spectroscopy bounds.

To demonstrate our approach, we select several representative values for the product of the coupling constants from the range $|g_1 g_2| \simeq 10^{-20}$-$10^{-18}$, which are approximately aligned with the constraints from precision spectroscopy. We also assume that the masses of the new scalar bosons are less than several 10 TeV so that they can be reached by a direct detection in future high-energy experiments. In addition, the new scalar bosons may induce proton or nuclear instability (see e.g., Ref. \cite{dev2024searches}). Implementing extra discrete symmetries in the related models could help avoid excessively rapid proton decay, as discussed in Ref. \cite{mohapatra1983spontaneous}, ensuring that the selected representative values remain consistent with the existing experimental limits on the proton lifetime $\tau_p \gtrsim 10^{31}$-$10^{33}$ yr \cite{navas2024review}.

Ref. \cite{goity1995bounds} provides an approximate formula for calculating the decay rate of the $nn \rightarrow VV$ process, which can be further simplified under quasi-free assumptions and by ignoring the nuclear binding energy and the Fermi motion as follows \cite{he2021eft,he2021b}:  
\begin{widetext}
\begin{equation}
\begin{split}
\Gamma_{nn \rightarrow VV} &\simeq \frac{\rho_{n}}{32 \pi S m_{n}^2} K(1,\xi,\xi)^{\frac{1}{2}} \overline{|\mathscr{M} (nn \rightarrow VV)|}^2, \\
&\simeq K(1,\xi,\xi)^{\frac{1}{2}} \rho_{n} N_{f} g_1^2 g_2^2  \frac{4m_n^4 - 4 m_n^2 M_V^2 + 3 M_V^4}{32 \pi S M_V^2 (4 m_n^2 - M_{\phi}^2)^2}.
\end{split}
\label{eqrate}
\end{equation}
\end{widetext}
Here, the squared amplitude is calculated by summing over all final spin configurations and averaging over all initial spin configurations:
\begin{equation}
\begin{split}
&\overline{|\mathscr{M} (nn \rightarrow VV)|}^2 \\
=&\frac{1}{4}g_1^2 g_2^2 M_V^2 g_{\mu \nu} g_{\alpha \beta} \big(-g^{\mu \alpha} + \frac{p_{3}^{\mu}p_{3}^{\alpha}}{M_V^2}\big) \big(-g^{\nu \beta} + \frac{p_{4}^{\nu}p_{4}^{\beta}}{M_V^2}\big) \\
&\times \frac{1}{\big[\big(p_1 +p_2\big)^2 -M_{\phi}^2\big]^2}\text{Tr} \left[ (\cancel{p_1}+m_n) (\cancel{p_2}+m_n)  \right] \\
\simeq & N_{f} g_1^2 g_2^2  \frac{ m_n^2(4m_n^4 - 4 m_n^2 M_V^2 + 3 M_V^4)}{M_V^2 (4 m_n^2 - M_{\phi}^2)^2}.
\end{split}
\label{amplitude}
\end{equation}
In Eq. (\ref{eqrate}), $K(x,y,z)$ denotes the Kallen triangle function and can be expressed as $K(x,y,z) \equiv x^2 + y^2 + z^2 -2xy-2yz-2zx$. The dimensionless parameter $\xi$ is given by $\xi \equiv M_{V}^2/(4 m_{n}^2)$ \cite{he2021eft}. The neutron number density is denoted by $\rho_{n}$. We also assume that the mass of the dark photon is very small and satisfies the relation: $M_{V} \ll m_{n}$. The parameters $S$ and $N_f$ denote the symmetry factor and the numerical factor, respectively. In our case, they take the values: $S=2$ and $N_f=2$.

The dineutron decay rate given in Eq. (\ref{eqrate}) was originally derived for low-mass atomic nuclei, especially for the $^{16}$O nucleus, via the effective field theory (EFT) approach with high dimensional operators \cite{he2021eft,he2021b}. The validity of Eq. (\ref{eqrate}) for the neutron-star matter is supported by the characteristic features of the EFT approach, where only the low-energy effective degrees of freedom are used to make the calculations tractable while the untractable degrees of freedom can be absorbed into the effective coupling parameters. Likewise, Eq. (\ref{eqrate}) is only a low-energy effective formula of the dineutron decay rate and the relevant difference in the effects of dense nuclear medium can be absorbed into the effective coupling parameters.

Although neutron stars may exhibit different nuclear environments and significantly higher central densities, their average densities are still comparable to the densities of atomic nuclei \cite{camenzind2007compact}. Due to the main features of the EFT approach, our evaluations are only valid up to the order of the magnitude. Therefore, it is reasonable to assume that the difference in the nuclear medium effects, such as the mean-field potential, Fermi momentum, etc., could only shift the dineutron decay rate mildly and would not qualitatively affect our final conclusions.

In fact, the physical properties of super-dense nuclear matter are still uncertain. Theoretical models of the neutron-star matter rely heavily on theoretical assumptions regarding high-density matter. Although the gravitational wave detection may provide indirect information on the internal structure and the EoS of the neutron star \cite{hinderer2010tidal}, there is a lack of direct experimental information on the neutron-star matter, making the corresponding parameter space poorly constrained. Moreover, the properties of low-mass nuclei are accessible in the terrestrial laboratories, whereas those of the neutron-star matter are not. To some extent, the properties of low-mass nuclei could help benchmark the properties of the neutron-star matter. Or at least, they could provide some insights and clues into the properties of the neutron-star matter. Unless the assumption is refuted by future experimental results, we consider Eq. (\ref{eqrate}) can reasonably be extended to the analysis of the neutron-star matter.

The dineutron decay rate in Eq. (\ref{eqrate}) is obtained based on the quasi-free assumptions. Here, the concept of dineutron does not refer to pairing, clustering, or resonant states as understood in nuclear physics (see e.g., Refs. \cite{ivanytskyi2019tetraneutron,pais2023influence,seth1991evidence,kisamori2016candidate}), but rather to two quasi-free neutrons participating in the $nn \rightarrow VV$ process in a neutron-dominated environment like the neutron star. When modeling scattering processes in the neutron star, we could treat neutrons as quasi-free particles and safely ignore their correlations. Treating neutrons as quasi-free particles in a dense, degenerate environment, such as nuclei and neutron stars, has been widely adopted when evaluating BNV effects in previous studies (see e.g. Refs. \cite{berezhiani2021neutron,berryman2022neutron,he2021eft,he2021b}). The correlation effects can be absorbed into density-dependent effective masses or coupling parameters (see e.g. Ref. \cite{akmal1998equation}). The discrepancy associated with the neglect of the correlation effects can be compensated by adjusting the effective masses and the coupling parameters.

In a physically more rigorous treatment, the neutron mass used in the analysis of the BNV processes inside neutron stars should not strictly be the free (bare) neutron mass $m_n$, but rather an effective mass $m_n^*$ \cite{berryman2024macroscopic}. The effective neutron mass $m_n^*$ characterizes the response of a neutron to external interactions and perturbations due to the effects of the dense nuclear medium. In fact, it captures the dynamical properties of the neutron inside neutron stars \cite{chamel2006effective}, such as transport properties, superfluid properties, thermodynamic properties, etc. The discrepancy between the free neutron mass $m_n$ and the effective neutron mass $m_n^*$ tends to be less than about 30\% in most cases \cite{jha2006neutron}. Since the EFT interaction mediating the $nn \rightarrow VV$ process comes from integrating out high-energy degrees of freedom, the dineutron decay can be effectively treated as a localized contact interaction and the medium effects only arise at the sub-leading order. This approximation is consistent with the treatment in Refs.\cite{he2021eft,he2021b}, where the medium effects such as Fermi motion and nuclear correlations are ignored. Since our evaluations are only valid up to the order of the magnitude, using the free neutron mass instead of the effective neutron mass in the medium would not qualitatively affect our conclusions either.

Furthermore, the values of $m_n^*$ depend heavily on the microscopic models and local densities \cite{stone2003nuclear,jha2006neutron,chamel2006effective} and vary widely in different references (see e.g. Refs. \cite{babaev2009unconventional,hutauruk2022effect}). In the study of the BNV processes inside neutron stars, approximating the effective neutron mass by the free neutron mass is a common simplification (see e.g. Ref. \cite{goldman2025effects}). The substitution of $m_n^*$ with $m_n$ tends to overestimate the mass loss of the neutron star and thus tends to make our bounds slightly conservative. Since the dineutron decay rate depends on both the neutron mass and the coupling constants, no inconsistencies would occur if we carefully adjust the values of the coupling constants when using the free neutron mass to evaluate the $nn \rightarrow VV$ decay rate.

As mentioned above, the effective neutron mass remains close to the free neutron mass (up to several tens of percent) in the calculation of the $nn \rightarrow VV$ process. In fact, the effective mass could significantly influence the EoS of the neutron-star matter and the properties of the neutron star \cite{jha2006neutron}. Since the effective mass is a key factor in the EoS, it could influence the $nn \rightarrow VV$ process indirectly via the EoS, but not via an explicit factor in the decay rate and red-shift formulae. In this regard, the effect of the effective mass can be partially and implicitly included in our calculations through the adopted EoS.

\section{Hydrostatic equilibrium of neutron stars}

We will first review the parameterization schemes for the EoS of neutron-star matter and explain the reasons for our choice of the EoS. Then, we will review the differential equations that describe the structure of neutron stars in hydrostatic equilibrium.

The EoS of the neutron star makes a connection between microscopic elementary particles and macroscopic celestial bodies, and is a key factor in the calculation of neutron-star properties \cite{oertel2017equations}. Since neutron stars tend to have very different physical properties from the surface to the core \cite{potekhin2015neutron,haensel2007neutron}, piecewise-type functions can be used to model their properties. In the $i$-th subdomain, the pressure $P(r)$ and energy density $\epsilon(r)$ can de defined by \cite{read2009constraints,ozel2009reconstructing}: 
\begin{align}
&\begin{aligned}
P(r) &\equiv& K_j \rho(r)^{b_j}, \nonumber
\end{aligned}\\
&\begin{aligned}
\epsilon(r) &\equiv& (1 + c_j) \rho(r) + \frac{K_j}{b_j -1} \rho(r)^{b_j},
\end{aligned}
\end{align}
with the coefficient $c_j$ determined by the continuity condition \cite{smith2012tolman}:
\begin{align}
&\begin{aligned}
c_0 &=& 0,
\end{aligned}\\
&\begin{aligned}
c_j &=& c_{j-1} + \frac{K_{j-1}}{b_{j-1} -1} \rho_{j}^{b_{j-1} -1} - \frac{K_{j}}{b_{j} -1} \rho_{j}^{b_{j} -1}.
\end{aligned}
\end{align}
Here, the subscript $j$ ($j=1,2,3,\ldots,N$) denotes each subdomains of the neutron star. $b_j$ is the adiabatic index, and $K_j$ is a normalization factor. The piecewise-polytropic parameterization scheme with three different adiabatic indices ($b_1$, $b_2$, and $b_3$) and a reference pressure $P_1$ could efficiently simulate the EoSs of neutron-star matter \cite{read2009constraints}. Since the validity of the parameterization scheme of the EoSs proposed in Ref. \cite{read2009constraints} has been demonstrated in several aspects \cite{takami2015spectral,lackey2015reconstructing}, we prefer such a scheme for our numerical simulation.

Since there is no direct experimental data on the interiors of the neutron star \cite{oertel2017equations}, the theoretical models of the neutron-star matter rely heavily on theoretical assumptions regarding the behavior of matter under extreme conditions. The parameter space of the theoretical models remains poorly constrained. As a result, the choice of EoSs for the neutron-star matter would inevitably cause systematic uncertainties in the numerical results, reflecting our limited knowledge of dense matter. Despite the uncertainties, a broad class of the EoSs tends to give rise to the results within the same order of magnitude, which are still adequate for capturing the main features of the BNV effects and for deriving the constraints on the parameter space \cite{berryman2022neutron}.

The maximum neutron-star mass predicted by an EoS can be used as a practical criterion for evaluating its feasibility \cite{ozel2016masses}. Observations show that some systems, such as PSR J0740+6620 \cite{fonseca2021refined} and GW190814 \cite{abbott2020gw190814}, are believed to contain neutron stars heavier than 2$M_\odot$. This suggests that the EoSs, which lead to the maximum neutron-star value below 2$M_\odot$, can be ruled out. Ref. \cite{biswas2022bayesian} summarized a more comprehensive list of EoSs that meet this criterion. In particular, the MPA1 EoS \cite{muther1987nuclear} has  special advantages in modeling the properties of the neutron star (see e.g. Ref. \cite{biswas2022bayesian,odintsov2023inflationary}). Given the above reasons, we prefer the MPA1 EoS for our calculations.

In our analysis, the neutron stars are assumed to be static and respect spherical symmetry. They can be described by the following metric \cite{hartle1967slowly,tolman1987relativity,kogut2018special}:
\begin{equation}
\begin{split}
ds^2 &= g_{\mu \nu} dx^{\mu} dx^{\nu} \\
     &= e^{2 \Phi(r)}dt^2 - \Big(1-  \frac{2GM}{r} \Big)^{-1}dr^2  -r^2 d\Omega^2,
\end{split}
\end{equation}
where $G$ denotes the gravitational constant, and the metric $d\Omega^2$ is defined by
\begin{equation}
d\Omega^2 \equiv d\theta^2 + \sin^2 \theta d\phi^2.
\end{equation}
Here, $\Phi(r)$ can be considered as the effective gravitational potential associated with the time component of the metric tensor $g_{\mu \nu}$ and can be evaluated by \cite{fattoyev2010sensitivity}
\begin{equation}
\begin{split}
\Phi(r) \equiv & \int^{R}_{r}\frac{r^{\prime}}{r^{\prime}-2GM(r^{\prime})} \Big[\frac{GM(r^{\prime})}{r^{\prime 2}} + 4 \pi G r^{\prime} P(r^{\prime})\Big]dr^{\prime}\\
               & - \frac{1}{2} \ln{\Bigl[1 - \frac{2 G M(R)}{R} \Bigr]}, \quad 0 < r^{\prime} \leq R.
\end{split}
\end{equation}

The structure of the neutron star in hydro-static equilibrium is determined by the Tolman-Oppenheimer-Volkoff (TOV) equations \cite{oppenheimer1939massive,tolman1939static}:
\begin{eqnarray}\label{TOV}
\frac{dP(r)}{dr} &=& - \frac{[\epsilon(r) + P(r)][M(r) + 4\pi r^3 P(r)]}{r[r-2 G M(r)]}, \nonumber \\
\frac{dM(r)}{dr} &=& 4\pi r^2 \epsilon(r), \nonumber \\
\frac{d\Phi(r)}{dr} &=& \frac{r}{r-2GM(r)} \Big[\frac{GM(r)}{r^2} + 4 \pi GrP(r)\Big],
\end{eqnarray}   
where $M(r)$ is the mass within the spherical volume of the radius $r$. $P(r)$ and $\epsilon(r)$ are the pressure and energy density, respectively. The Fourth-Order Runge-Kutta (RK4) \cite{press2007numerical} method can be used to numerically solve the TOV equations under the following boundary conditions: $\rho(0) \equiv \rho_c$, $P(R)\equiv 0$, where $R$ and $\rho_c$ are the radius and the central density of the neutron star, respectively.

In the above-mentioned notations, the total number of neutrons contained in a neutron star is related to the nucleon-number density by \cite{glendenning2012compact}
\begin{equation}
N_n \equiv 4 \pi \int^R_0 \frac{r^2 \rho_{n}(r)}{\sqrt{1-  \frac{2GM}{r} }} dr \simeq 4 \pi \int^R_0 \frac{r^2 X_n\rho_{a}(r)}{\sqrt{1-  \frac{2GM}{r} }} dr,
\end{equation}
where $\rho_{a}(r)$ is the nucleon-number density and is related to the neutron number density by $\rho_{n}(r) \simeq X_n\rho_{a}(r)$. The parameter $X_n$ is the neutron fraction and is assumed to have the value $X_n \simeq 0.89$ \cite{berryman2022neutron}.

Since the neutron star would eventually lose mass due to the $nn \rightarrow VV$ process, it, in principle, would not maintain hydrostatic equilibrium. Following Ref. \cite{berryman2022neutron}, the TOV equations can still be valid even in the presence of the $nn \rightarrow VV$ process, because the dineutron decay timescale is so long enough that the neutron star evolves slowly enough to maintain quasi-equilibrium. We will explore the detectable consequences of the $nn \rightarrow VV$ process on the main properties of neutron stars by numerically solving the TOV equation in the following section.

\section{Observable effects \label{suba}}

\begin{figure}[t] 
\centering
\includegraphics[scale=0.99,width=0.99\linewidth]{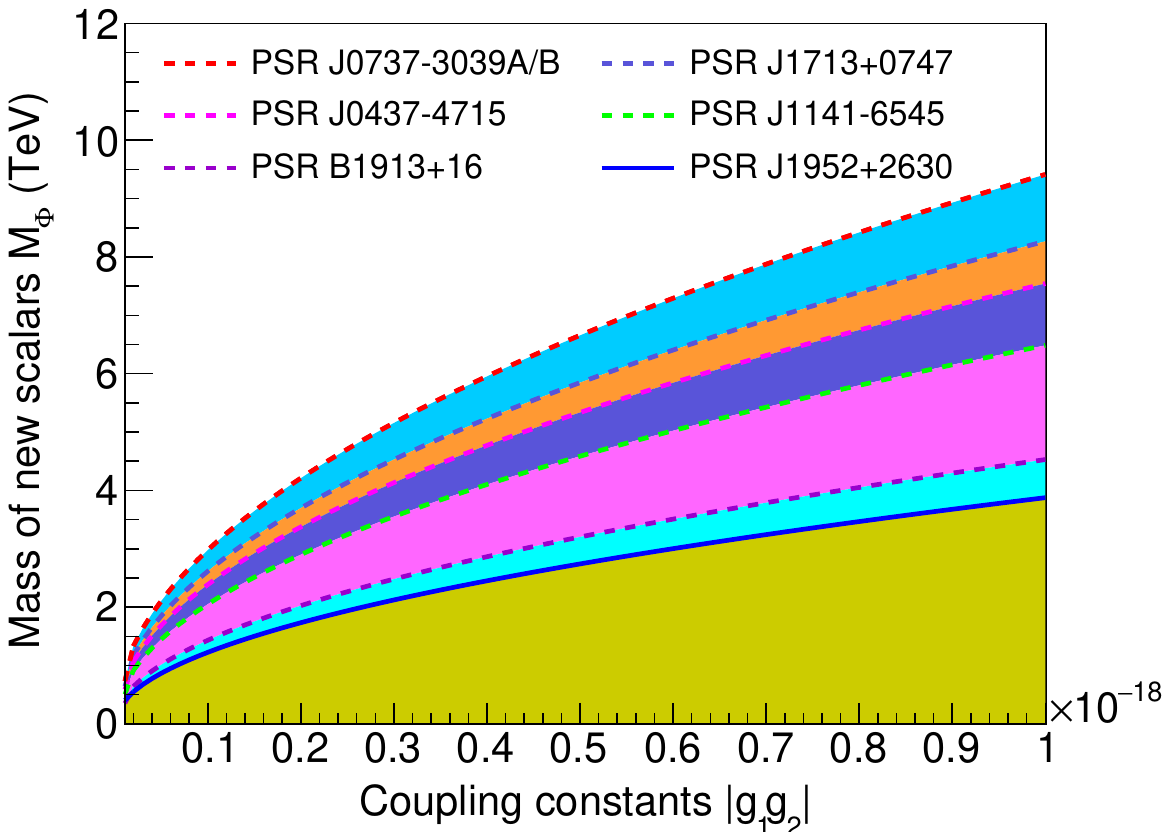}
\caption{Lower bounds on the scalar mediator mass $M_\phi$ as a function of the coupling product $|g_1 g_2|$, derived from the residual changes in the orbital period observed in various binary pulsar systems. Each curve corresponds to a specific binary system, with the constraints reflecting the sensitivity of the inferred mass loss due to the $nn \to VV$ process. (Color online)
}
\label{fig02}
\end{figure}

\subsection{Orbital-period variation of binary systems \label{subb}}

\begin{table*}[t]
\caption{Inferred lower bounds on the mass of the new scalar bosons ($M_{\phi}$) based on the residual mass losses and orbital-period anomalies observed in the set of well-measured binary pulsar systems in two coupling scenarios: $|g_1 g_2| = 10^{-19}$ and $10^{-18}$.}
\begin{ruledtabular}
\begin{tabular}{l|cccccccc}                                         
\diagbox{Binary sys.}{Parameter}& $M_1$ ($M_{\odot}$) & $M_2$ ($M_{\odot}$) & $|\Dot{P}/P|_{\text{BNV}}$& $|\Dot{M}/M|_{\text{BNV}}$& $^k$$M_{\phi}$ (TeV) & $^l$$M_{\phi}$ (TeV)\\\hline           
 PSR J0437-4715 &  $^a$1.76 & $^a$0.254  &  $-$  & $^g$$1.6 \times 10^{-11}$&3.88 &1.23  \\
 PSR B1913+16   &   $^b$1.438 & $^b$1.390   &  $-$  & $^i$$6.5 \times 10^{-13}$&8.28  &2.61  \\
                &           &           &  $^j$$5.2 \times 10^{-12}$   & $-$  & 5.85  &  1.85 \\
 PSR J1952+2630 & $^c$1.35  & $^c$0.93-1.48  &  $-$  & $^g$$7 \times 10^{-12}$&4.53 &1.43  \\
 PSR J0737-3039A/B &  $^d$1.338185 &  $^d$1.248868 &  $^h$$7.3 \times 10^{-13}$  & $-$&9.41 &2.98  \\
 PSR J1713+0747 &  $^e$1.33 & $^e$0.29  &  $^h$$1.8 \times 10^{-12}$  & $-$&7.54 &2.39  \\
 PSR J1141-6545&  $^f$1.27 & $^f$1.02  &  $-$  & $^g$$1.6 \times 10^{-12}$&6.48 &2.05  \\
\end{tabular}
\end{ruledtabular}
\begin{tablenotes}
\footnotesize
\centering
\item[\emph{a}]{
\begin{flushleft}
$^a$ Ref. \cite{verbiest2008precision}, $^b$ Ref. \cite{weisberg2016relativistic}, $^c$ Ref. \cite{lazarus2014timing}, $^d$ Ref. \cite{kramer2021strong}, $^e$ Ref. \cite{zhu2019tests}, $^f$ Ref. \cite{bhat2008gravitational}, $^g$ Ref. \cite{goldman2019bounds}, $^h$ Ref. \cite{berryman2022neutron}, $^i$ Ref. \cite{berezhiani2021neutron}, $^j$ Ref. \cite{goldman2025effects}; $^k$ These bounds are obtained using the coupling constant $|g_1 g_2| = 10^{-19}$; $^l$ These bounds are obtained using the coupling constant $|g_1 g_2| =  10^{-18}$.
\end{flushleft}
}
\end{tablenotes}
\centering
\label{tabone}
\end{table*}

We first analyze the effects of the $nn \rightarrow VV$ process on the orbital-period variation of binary pulsar systems. The potential decay of dineutrons into dark photons ($nn \rightarrow VV$) inside a neutron star implies a gradual loss of the neutron-star mass. Since neutron stars are often contained in tightly bound binary pulsar systems, such a process may give rise to measurable astrophysical effects, such as the possible orbital-period anomalies. As mentioned in the Introduction section, to account for the matter-antimatter asymmetry, many new-physics models predicted that the BNV processes might exist. However, so far, such processes have not been confirmed, suggesting that their magnitude would be very small. Binary pulsar systems that contain neutron stars are considered to be precise astrophysical laboratories, where the BNV processes may occur with an expected more significant rate.

In the binary pulsar systems, the changes in the orbital period can originate from numerous possible factors (see e.g. Refs. \cite{lorimer2005handbook,verbiest2008precision}), such as galactic motions \cite{lorimer2005handbook}, gravitational wave emission \cite{weisberg2016relativistic}, kinematic effects \cite{shklovskii1970possible}, exotic factors including BNV effects \cite{berezhiani2021neutron,berryman2022neutron,goldman2025effects}, etc. In other words, if the BNV processes occurred, they could have extremely low rates or else would have already caused noticeable changes in the orbital periods of binary systems. To begin with, the orbital period in binary pulsar systems can be measured with very high precision. From theoretical aspects, usually there is little room for residual changes in the orbital period after accounting for all known effects. Any deviation beyond theoretical expectations could imply new physics beyond the SM. The combined analysis based on theoretical modeling and comparison with high-precision observational data can put tight bounds on the BNV parameters. After deducting all known contributions, some residual changes or potential anomalies are not yet conclusively understood in several well-studied binary pulsar systems \cite{goldman2019bounds,berezhiani2021neutron,berryman2022neutron,goldman2025effects}. Following previous studies \cite{goldman2019bounds,berezhiani2021neutron,berryman2022neutron,goldman2025effects}, we assume that the $nn \rightarrow VV$ process is the dominant contribution to the orbital-period anomalies of the binary pulsar systems.

To quantify the orbital-period variation, the Jeans relation \cite{jeans1924cosmogonic} can be adopted to describe the response of the orbital period of the system to the slow mass loss. The fractional rate of the orbital-period variation of the binary pulsar systems can be approximated by

\begin{equation}
\frac{\Dot{P}_b}{P_b} \simeq -2 \frac{\Dot{M}}{M},
\end{equation}
where $M$ and $\Dot{M}$ denote the total mass and the mass-loss rate, respectively. $P_b$ and $\Dot{P}_b$ denote the orbital period and the rate of the orbital-period variation, respectively.

Within our scenario, the mass-loss rate arises as a result of the emission of dark photons through the $\mathcal{B}$-violating process. The total mass loss over time can be determined by integrating the energy depletion rate across the neutron star interior, taking into account the local matter density and the spatial dependence of the decay rate.

Assuming a spherically symmetric configuration and using the piecewise-polytropic EoS to model the structure of the neutron star, the total mass loss per unit time can be approximated as:
\begin{equation}
\begin{split}
\Dot{M} &\equiv \frac{d}{dt}\int_{0}^{R(t)} 4\pi r^2 \epsilon(r, t) dr\\
& \simeq \sum_{i}\int_{{\Delta V}_i} 4\pi r^2 \Big[ (1 + c_i) \rho + \frac{K_ib_i \rho^{b_i}}{b_i -1}\Big] \Gamma_{nn \rightarrow VV} dr,\\        
\end{split}
\label{massint}
\end{equation}
with the local dineutron decay rate approximately given by
\begin{equation}
\begin{split}
\Gamma_{nn \rightarrow VV} = \Big\vert \frac{\Dot{\rho}_n}{\rho_n}\Big\vert \simeq \Big\vert \frac{\Dot{\rho}_a}{\rho_a} \Big\vert \simeq  \Big\vert \frac{\Dot{\rho}}{\rho} \Big\vert.
\end{split}  
\label{approx}
\end{equation}
In Eq. (\ref{massint}), the index $i$ runs over every inner subdomains of the neutron star, and $\rho$ is the mass density. In Eq. (\ref{approx}), we have assumed that the temporal change in the neutron fraction is negligible (i.e. $\Dot{X}_n \simeq 0$) and the nucleon-number density $\rho_a$ is approximately proportional to the mass density $\rho$. This assumption follows previous studies (see e.g., Ref. \cite{berryman2022neutron}) where the BNV timescale is assumed to be much smaller than both the weak interaction timescale and the hydrodynamic adjustment timescale, so that the temporal change of the neutron fraction can be ignored. Similarly, we have assumed that the dineutron decay rate only faintly depends on neutron energy, and assumed that the nuclear medium and Fermi-motion effects in neutron stars are of a similar magnitude to those in atomic nuclei. Hence, Eq. (\ref{eqrate}) is equally applicable to the neutron-star matter under these assumptions.

The dineutron decay rate is sensitive to the coupling constants ($|g_1 g_2|$) and the mass scale of the mediating scalar bosons ($M_{\phi}$). The choice of the value of the coupling constants $|g_1 g_2|$ can benefit from the studies of isotopic shifts. Specifically, the bounds on similar coupling constants $|g_N g_e|$ from the studies of atomic and molecular transitions can be used as a benchmark reference value for the coupling constants $|g_1 g_2|$. In this case, the study of the $nn \rightarrow VV$ process may provide a cross check between particle physics experiments and atomic spectroscopy methods. Adopting representative values guided by atomic spectroscopy constraints, we assume that the plausible ranges for the product of coupling constants $|g_N g_e|$ and mediator mass $M_{\phi}$ could respectively be $|g_1 g_2| \sim 10^{-20}$-$10^{-18}$ and $M_{\phi} \lesssim 10$ TeV, which remain compatible with experimental bounds. This is a conservative assumption because the values we choose are generally lower than the upper bounds on $|g_N g_e|$ which are imposed by the precision atomic and molecular spectroscopy. Since the light (e.g. sub-MeV) dark photons as dark matter candidates have attracted distinctive interests both cosmologically and astrophysically \cite{cyncynates2024experimental}, we also choose a typical value for the mass of the dark photon  (i.e. $M_V \equiv 1$ keV), which is generally consistent with the derived bounds from experiments \cite{caputo2021dark}, to demonstrate the effects of the $nn \rightarrow VV$ process.

\begin{figure}[t] 
\centering
\includegraphics[scale=0.99,width=0.99\linewidth]{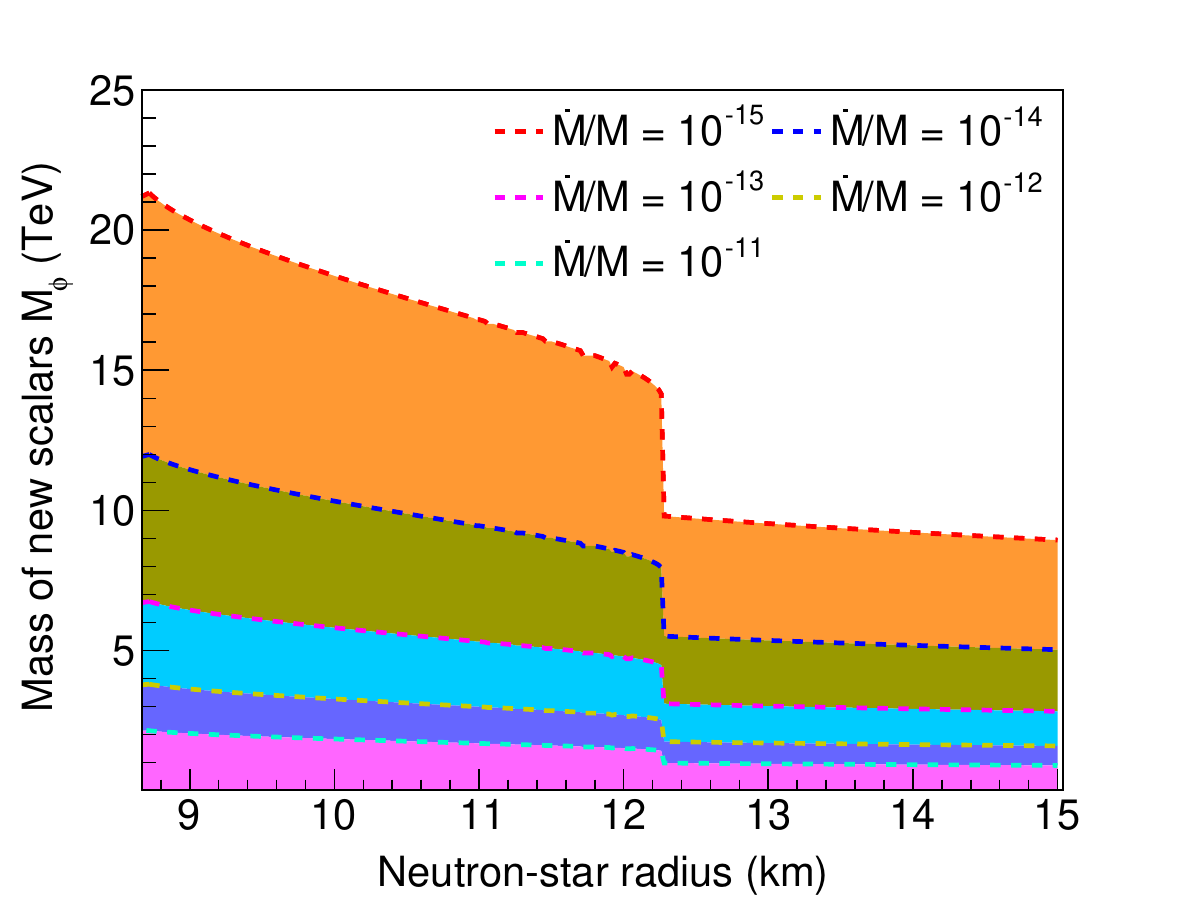}
\caption{Inferred lower bounds for $|g_1 g_2| = 10^{-19}$ on the scalar boson mass $M_\phi$ as a function of the neutron-star radius, assuming that the mass loss is dominated by the $nn \rightarrow VV$ process. (Color online)
}
\label{fig03}
\end{figure}

\begin{figure}[t] 
\centering
\includegraphics[scale=0.99,width=0.99\linewidth]{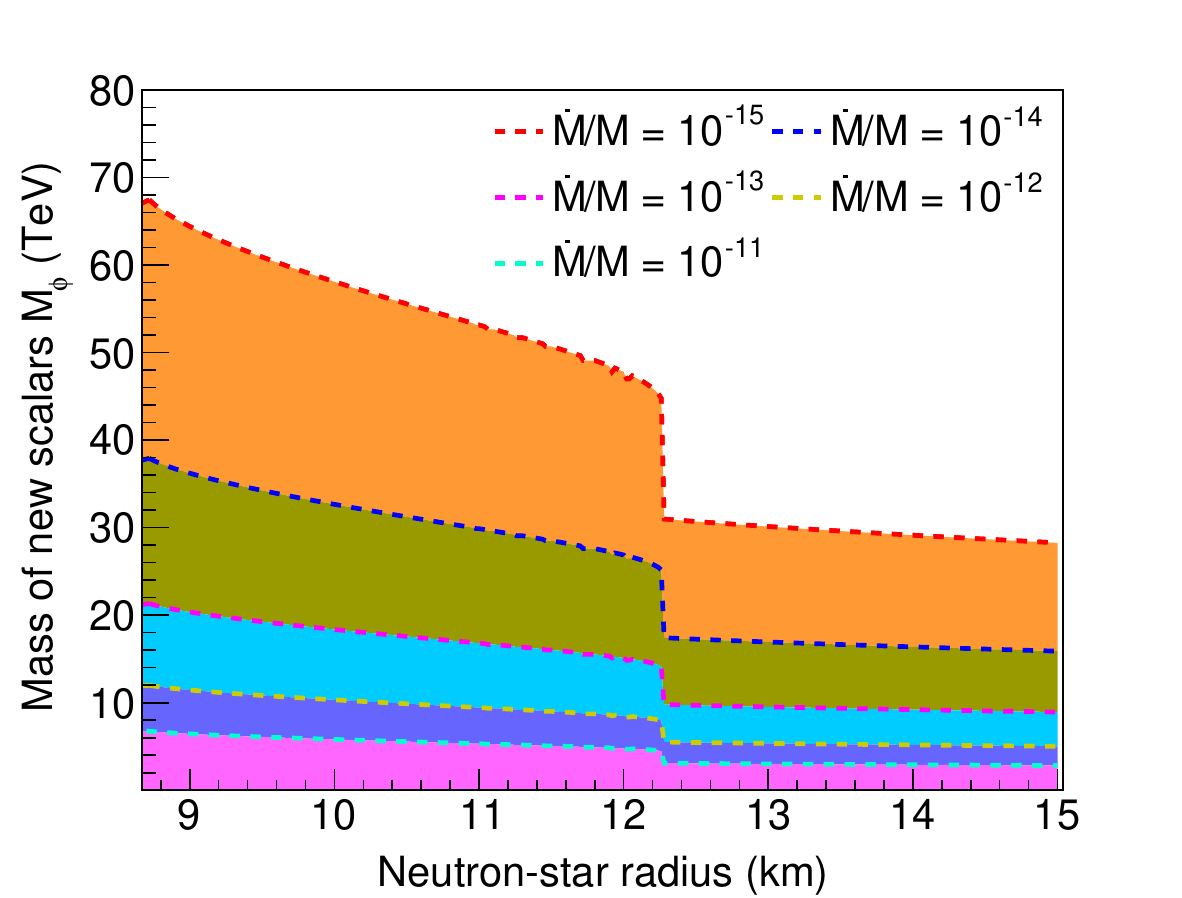}
\caption{Inferred lower bounds for $|g_1 g_2| = 10^{-18}$ on the scalar boson mass $M_\phi$ as a function of the neutron-star radius, assuming that the mass loss is dominated by the $nn \rightarrow VV$ process. (Color online)
}
\label{fig04}
\end{figure}

Table \ref{tabone} summarizes some well–characterized binary pulsar systems, including PSR J0437-4715 \cite{johnston1993discovery}, PSR B1913+16 \cite{hulse1975discovery}, PSR J1952+2630 \cite{knispel2011arecibo}, PSR J0737-3039A/B \cite{burgay2003increased}, PSR J1713+0747 \cite{foster1993new}, PSR J1141-6545 \cite{kaspi2000discovery}. 
The values $7.3 \times 10^{-13}$ yr $^{-1}$ \cite{berryman2022neutron}, $1.8 \times 10^{-12}$ yr $^{-1}$ \cite{berryman2022neutron}, and $5.2 \times 10^{-12}$ yr $^{-1}$ \cite{goldman2025effects} correspond to the residual changes in the orbital period for the binary systems PSR J0737-3039A/B, PSR J1713+0747, and PSR B1913+16, respectively. The values $1.6 \times 10^{-11}$ yr $^{-1}$ \cite{goldman2019bounds}, $6.5 \times 10^{-13}$ yr $^{-1}$ \cite{berezhiani2021neutron}, $7 \times 10^{-12}$ yr $^{-1}$ \cite{goldman2019bounds}, $1.6 \times 10^{-12}$ yr $^{-1}$ \cite{goldman2019bounds} correspond to the residual changes in the mass loss for the binary systems PSR J0437-4715, PSR B1913+16, PSR J1952+2630, and PSR J1141-6545, respectively.
Furthermore, we also translate the above residual changes into lower limits on the mediator mass for two benchmark products of couplings, $|g_1 g_2| \equiv 10^{-19}$ and $10^{-18}$. For the tighter coupling, the bounds span roughly in the range 4-9 TeV, whereas the coupling with an order‑of‑magnitude weaker relaxes them to the range 1-3 TeV, approximately. Among such binary systems, the double-neutron‑star system PSR J0737-3039A/B \cite{burgay2003increased}  yields the most stringent bounds and is followed by another double-neutron‑star system PSR B1913+16 \cite{weisberg2016relativistic} because the $nn \rightarrow VV$ process could occur in both components. The remaining systems are neutron star–white dwarf binaries \cite{verbiest2008precision,kaspi2000discovery,foster1993new,knispel2011arecibo} and contain only a single dark-photon emitter, making the corresponding bounds relatively less competitive.
All the limits are close to or already exceed the current direct‑search bounds at the LHC experiments (i.e., CMS \cite{cms2022searches1,cms2022search2} and ATLAS \cite{atlas2021search1,atlas2022search2}), but are expected to remain accessible to future collider upgrades or dedicated fixed‑target experiments.

Figure \ref{fig02} summarizes the inferred lower limits on the mass scale of the mediating scalar bosons ($M_{\phi}$), plotted as a function of the coupling constants ($|g_1 g_2|$). These limits are obtained by requiring that the mass loss in neutron stars caused by the $nn \rightarrow VV$ process does not exceed the residual mass loss or orbital-period anomalies observed in the set of well-measured binary pulsar systems presented in Tab. \ref{tabone}. As anticipated, smaller couplings require comparatively lighter scalar mediators to be consistent with the data, while larger couplings permit higher bounds on the mediator mass $M_{\phi}$. The variation between systems could imply a good accuracy of the derived constraints as well as a high precision of the astronomical observables, such as the orbital period and the total mass.

Figures \ref{fig03} and \ref{fig04} show the derived lower bounds on $M_{\phi}$ versus the neutron‑star radius for the typical assumed mass‑loss rates $|\Dot{M}/M|_{\text{BNV}}$ in the range from $10^{-15}$ to $10^{-11}$ yr$^{-1}$. Compact stars tend to tighten the constraints since a smaller radius more likely indicates a greater central density and thus a larger dineutron-decay rate. The curves drop quickly around $\sim$12 km but are nearly flat for less compact configurations. When the density drops below a threshold value, further decrease in the density barely alters the bound, as seen by the weak radius-dependence at large $R$. The same qualitative behavior can be identified in both figures. Such couplings uniformly increase the bounds by a few times when the coupling constants are increased by one order of magnitude.

\begin{figure}[t] 
\centering
\includegraphics[scale=0.99,width=0.99\linewidth]{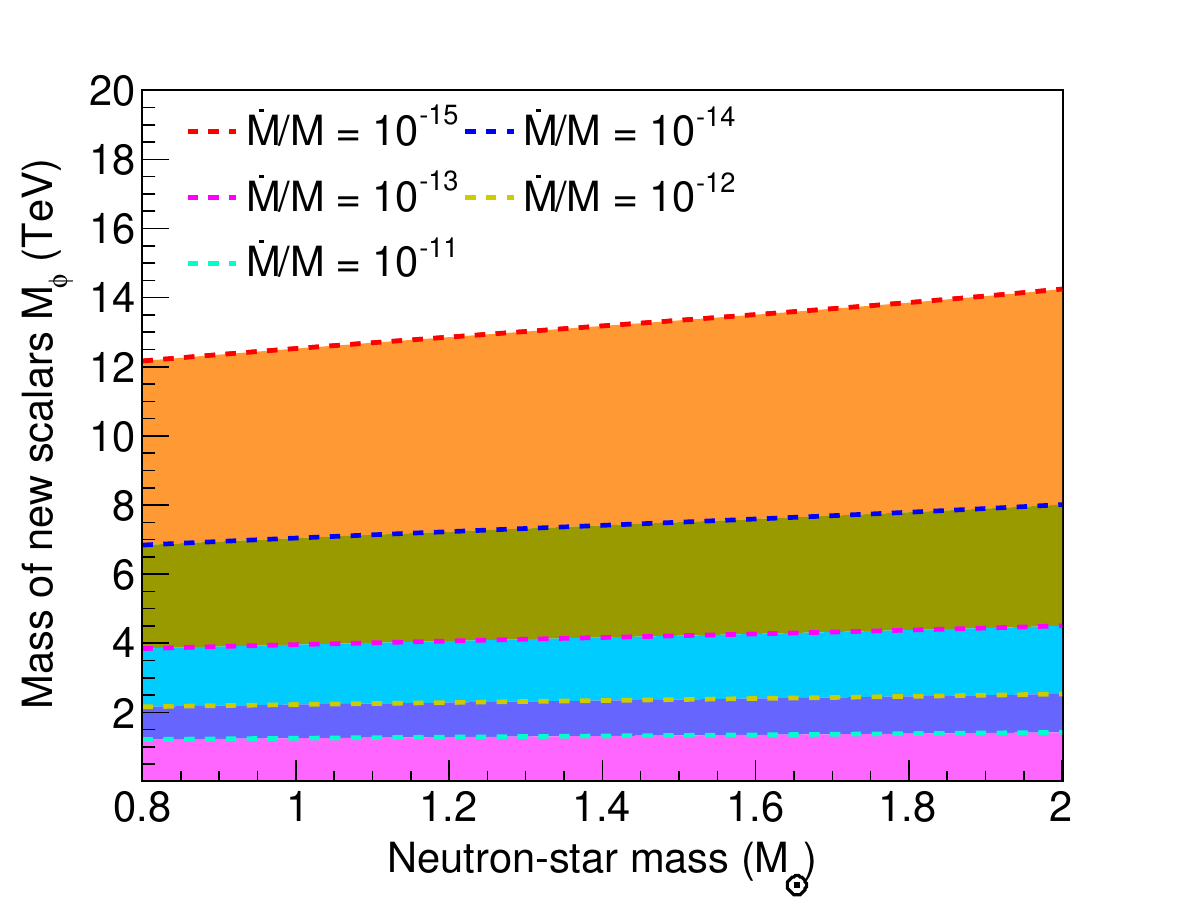}
\caption{Inferred lower bounds for $|g_1 g_2| = 10^{-19}$ on the scalar boson mass $M_\phi$ as a function of the neutron-star mass, assuming that the mass loss is dominated by the $nn \rightarrow VV$ process. (Color online)
}
\label{fig05}
\end{figure}

\begin{figure}[t] 
\centering
\includegraphics[scale=0.99,width=0.99\linewidth]{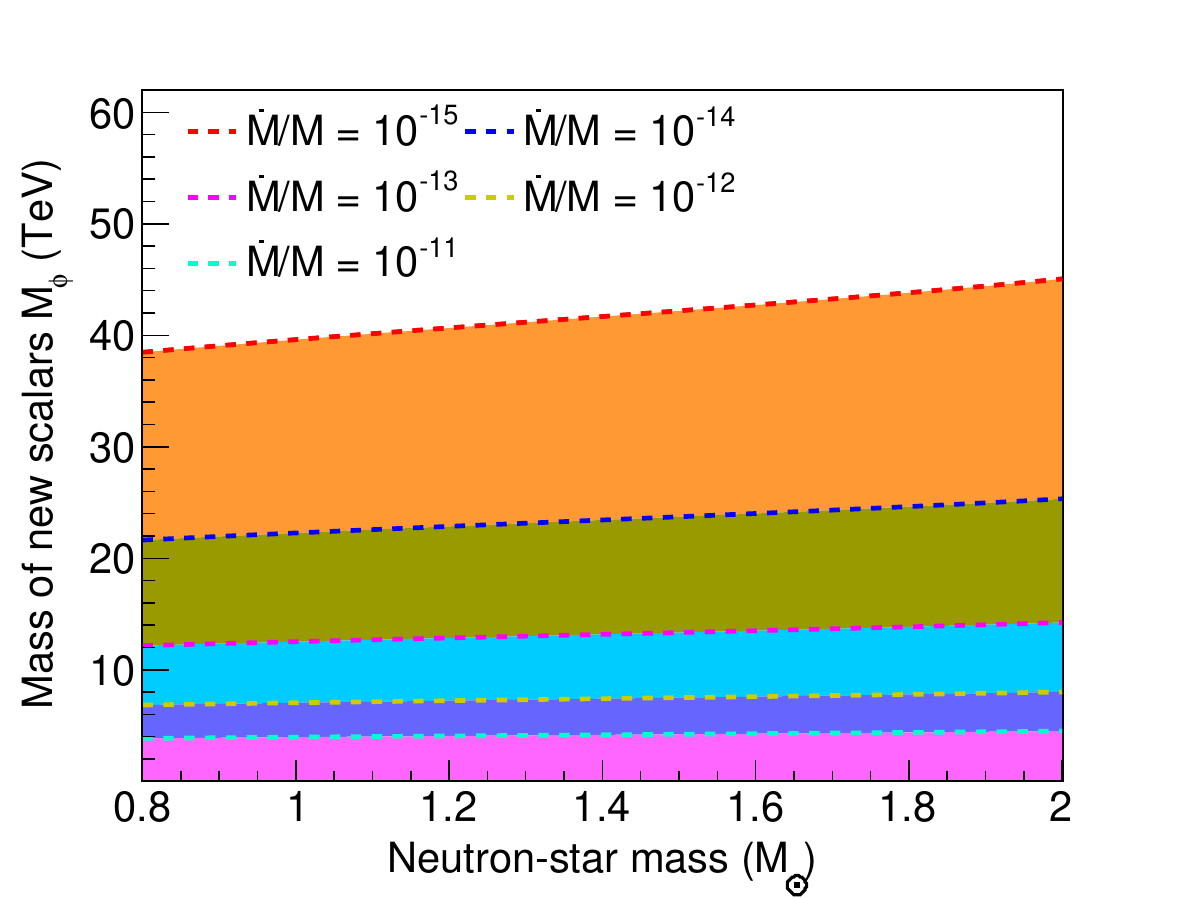}
\caption{Inferred lower bounds for $|g_1 g_2| = 10^{-18}$ on the scalar boson mass $M_\phi$ as a function of the neutron-star mass, assuming that the mass loss is dominated by the $nn \rightarrow VV$ process. (Color online)
 }
\label{fig06}
\end{figure}

Figures \ref{fig05} and \ref{fig06} show the derived lower bounds on $M_{\phi}$ versus the neutron-star mass for the typical assumed mass‑loss rates. An opposite increase trend can be seen in both figures, comparing with Figures \ref{fig03} and \ref{fig04}. From 1$M_{\odot}$ to 2$M_{\odot}$, the derived bounds only increase by about $\sim$10-20\%. Heavier neutron stars with smaller radii tend to give higher bounds because the $nn \rightarrow VV$ process could be accelerated by the deeper potential wells, which confine more neutrons with higher density.

Earlier investigations have indicated that the orbital-period anomalies of binary systems typically fall within the range of approximately $10^{-13}$ to $10^{-11}$ yr$^{-1}$ (see e.g. Refs. \cite{goldman2019bounds,berezhiani2021neutron,berryman2022neutron,goldman2025effects}). As observational technologies continue to develop, future improvements in measurement precision and modeling are expected to reduce the uncertainty in these residual changes, potentially reaching the $10^{-14}$ yr$^{-1}$ or even $10^{-15}$ yr$^{-1}$ level. Such an enhanced sensitivity would tighten the constraints on the feasible parameter space of new physics beyond the SM. Crucially, if these residual changes persist and cannot be accounted for within the SM framework, they could serve as compelling evidence for the existence of new physics.

Previous studies show that the BNV process would give rise to the gradual mass-loss of a neutron star and thus may affect the collapse thresholds or stability of the neutron star \cite{berryman2022neutron}, especially for configurations near the TOV limit (see e.g. Ref. \cite{haensel2007neutron} and the references therein) or the minimum critical limit \cite{colpi1989exploding,blinnikov1990explosion} allowed by the EoS. 
Likewise, the $nn \rightarrow VV$ process may have non-trivial effects on the secular gravitational stability of the neutron star. On the one hand, this process might postpone the collapse of a super-heavy neutron star into a black hole by shifting the neutron star away from instability through a slow loss of mass. On the other hand, this process might also lead to the explosion of the neutron star near the minimum mass \cite{colpi1989exploding,blinnikov1990explosion,berryman2022neutron}.

The leading-order phase evolution of the gravitational wave from a compact binary system is closely related to the chirp mass $\mathcal{M} \equiv (M_1 M_2)^{3/5}(M_1 + M_2)^{-1/5}$ \cite{cutler1994gravitational}. Here, $M_1$ and $M_2$ are the masses of the two components contained in the binary system. The $nn \rightarrow VV$ process could give rise to continuous mass loss from neutron stars over long timescales and thus might lead to a shift in the gravitational wave spectrum. Furthermore, the long-term BNV effects arising from the $nn \rightarrow VV$ process may affect the tiny secular change of the tidal deformability \cite{hinderer2010tidal} or the mass quadrupole moment \cite{thorne1980multipole} of the neutron star, and then may influence the signatures of the gravitational wave signals in the LIGO or Virgo detectors \cite{abbott2016observation}.

\begin{figure*}[t] 
\centering
\includegraphics[scale=0.99,width=0.99\linewidth]{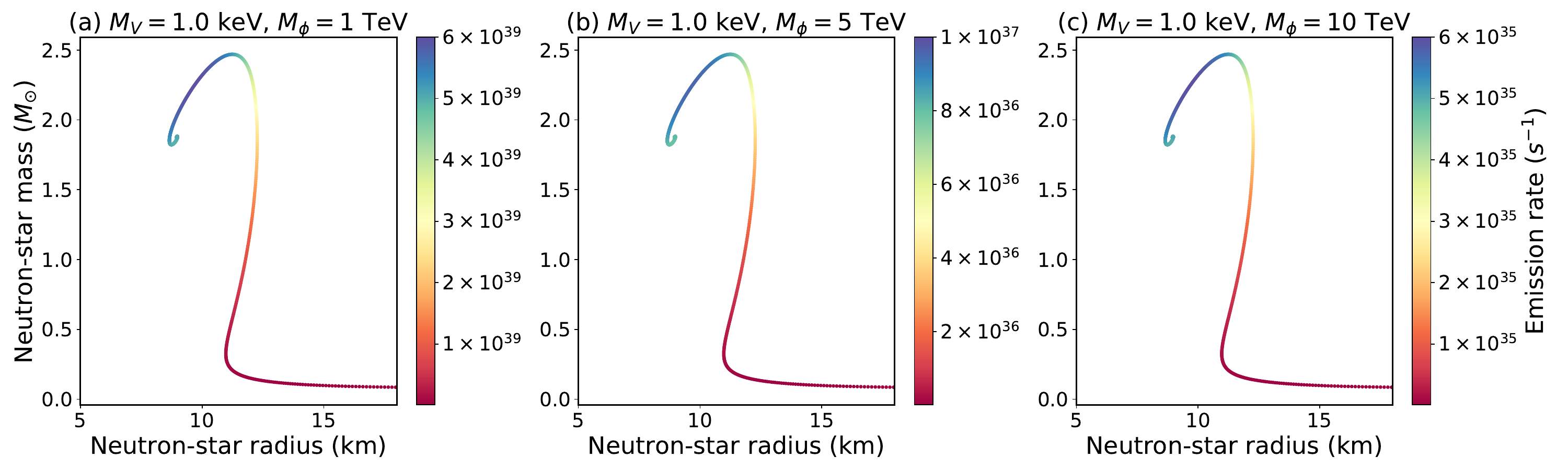}
\caption{Estimated dark-photon emission rate plotted in the mass-radius plane of the neutron star for several representative values of the coupling product $|g_1 g_2|$ and the scalar mediator mass $M_\phi$: (a) $|g_1 g_2| \equiv 1 \times 10^{-19}$, $M_{\phi} \equiv 1$ TeV; (b) $|g_1 g_2| \equiv 1 \times 10^{-19}$, $M_{\phi} \equiv 5$ TeV; (a) $|g_1 g_2| \equiv 1 \times 10^{-19}$, $M_{\phi} \equiv 10$ TeV. (Color online)
}
\label{fig07}
\end{figure*}

\begin{figure*}[t] 
\centering
\includegraphics[scale=0.99,width=0.99\linewidth]{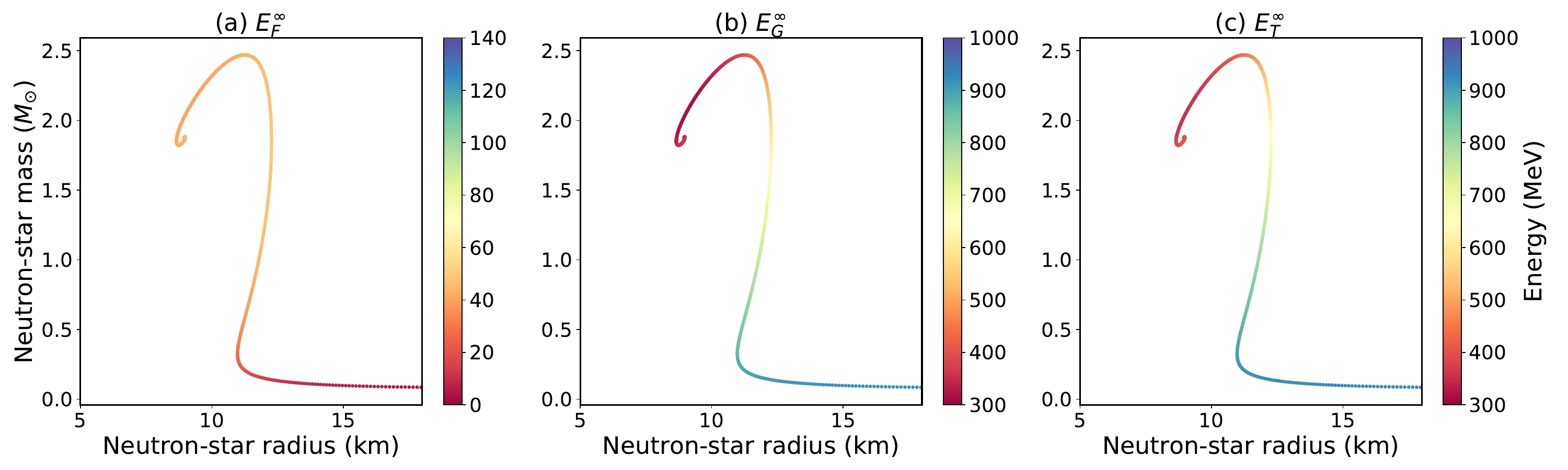}
\caption{Red-shifted energy spectrum of dark photons emitted from the neutron star as a function of the neutron-star mass and radius for several representative values of the coupling product $|g_1 g_2| \equiv 1 \times 10^{-19}$ and the scalar mediator mass $M_\phi$: (a) The red-shifted average Fermi energy per particle $E_{F}^{\infty}$; (b) The red-shifted gravitational binding energy of the rest mass per particle $E_{G}^{\infty}$; (c) The red-shifted total energy per particle $E_{T}^{\infty}$. (Color online)
}
\label{fig08}
\end{figure*}

To verify that our model aligns with current experimental data, we examined whether the presence of the $nn \rightarrow VV$ process is compatible with the observed lifetime of atomic nuclei. The simplest way to guarantee this compatibility is to assume that the $nn \rightarrow VV$ process is suppressed in normal nuclear environments and only becomes active above a specific critical neutron number density (see e.g. Ref. \cite{berryman2022neutron}). Similarly, the $nn \rightarrow VV$ process would be allowed in neutron stars but would be inactive in ordinary matter if this critical density were between that of atomic nuclei and that of neutron-star interiors.

Alternatively, taking into account the $nn \rightarrow VV$ process, the partial lifetime of the atomic nuclei can be directly calculated and compared to the current experimental limits. The experimental limits were reported for some low-mass nuclei, such as $^{12}$C ($T_{nn \rightarrow \text{inv.}} \gtrsim 1.4 \times 10^{30}$ yr \cite{araki2006search}), $^{16}$O ($T_{nn \rightarrow \text{inv.}} \gtrsim 1.5 \times 10^{28}$ yr \cite{allega2022improved}), etc. More experimental limits are summarized and discussed in Ref. \cite{heeck2020inclusive}. In order to illustrate the consistency, we take a spherical $^{12}$C nucleus of the charge radius $r_C = 2.4702$ fm \cite{angeli2013table} as an example to estimate its partial lifetime. Furthermore, we conservatively adopt a scalar boson mass $M_{\phi}$ close to the TeV scale and a coupling constant $|g_1 g_2| \simeq 10^{-18}$. The estimated limit on the partial life time of the $^{12}$C nucleus is $T_{nn \rightarrow VV}(^{12}\text{C}) \gtrsim 2 \times 10^{44}$ yr. Since such a value exceeds the experimental limits by many orders of magnitude, it suggests that, even for moderate values of scalar mass, the model remains consistent with the experimental constraints on the stability of nuclei.

\subsection{Emission of dark photons}

\begin{figure}[t] 
\centering
\includegraphics[scale=0.99,width=0.99\linewidth]{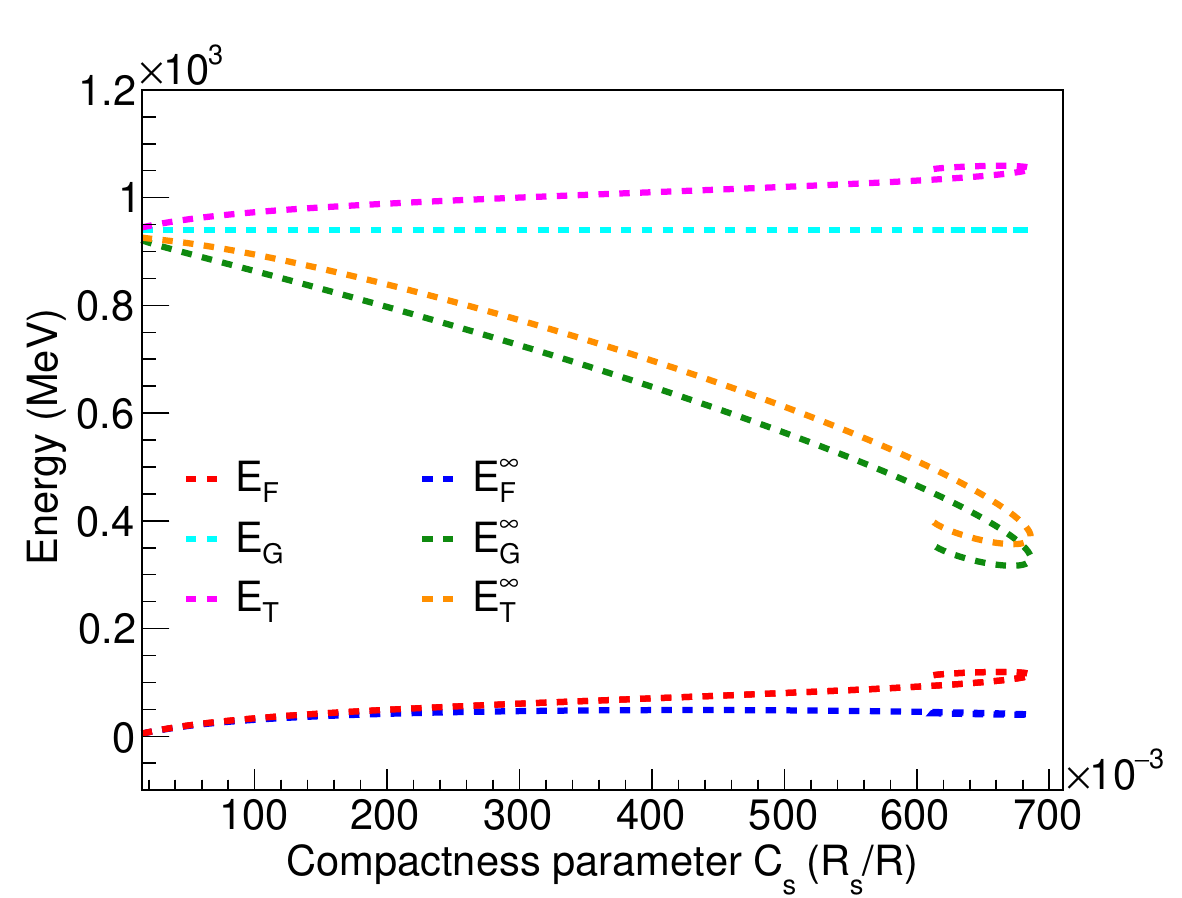}
\caption{Relationship between the red-shifted average energies ($E_{F}^{\infty}$, $E_{G}^{\infty}$, and $E_{T}^{\infty}$) of dark photons and the compactness parameter ($C_s \equiv R_s/R$) of the neutron star. (Color online)
}
\label{fig09}
\end{figure}

\begin{figure}[t] 
\centering
\includegraphics[scale=0.99,width=0.99\linewidth]{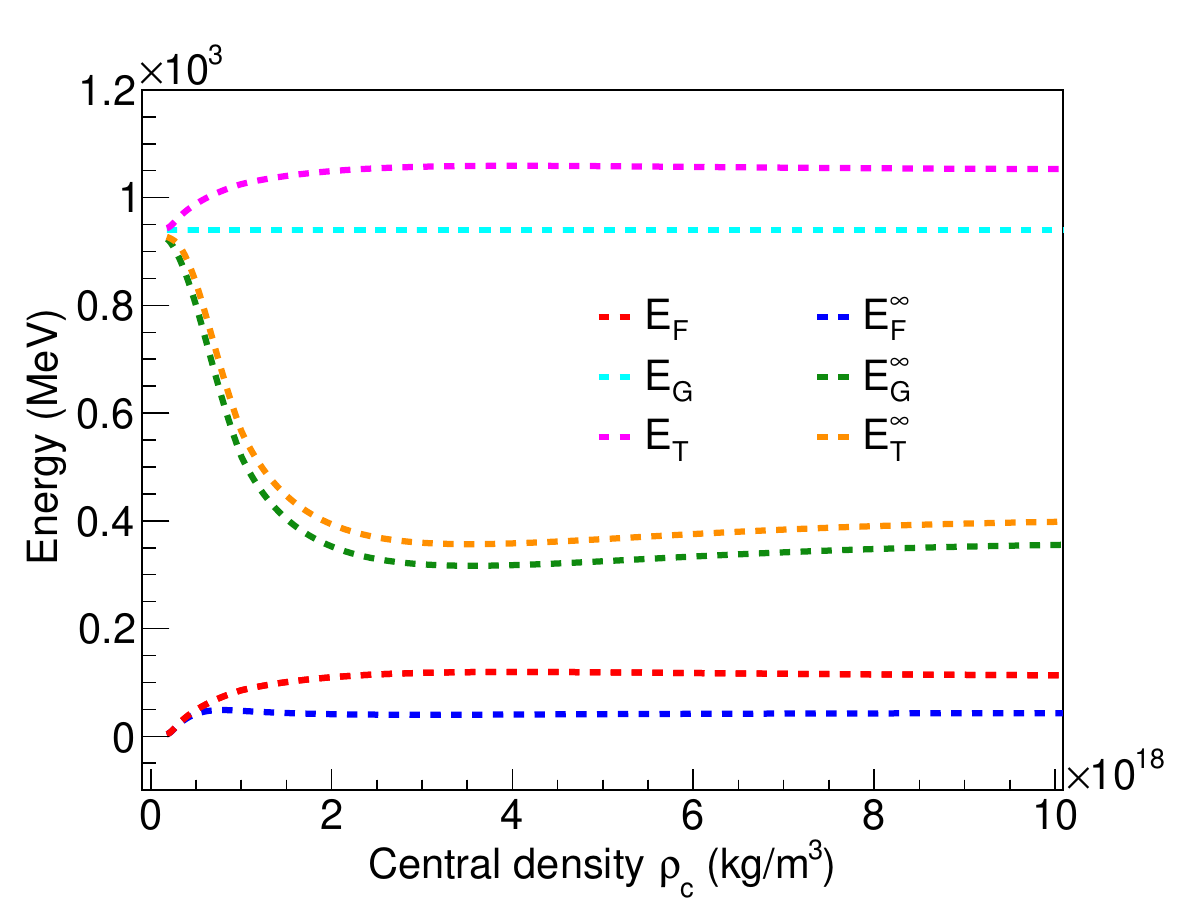}
\caption{Relationship between the red-shifted average energies ($E_{F}^{\infty}$, $E_{G}^{\infty}$, and $E_{T}^{\infty}$) of dark photons and the central density $\rho_c$ of the neutron star (Color online)
 }
\label{fig10}
\end{figure}

The emission of dark photons from the neutron stars would be the most direct physical consequence of the $nn \rightarrow VV$ process. Based on our assumptions, the dark photons ($V$) could be considered as dark-matter candidates (see e.g. Ref. \cite{fabbrichesi2021physics}) and hardly interact with the ordinary matter. They could move almost freely in the interior of the neutron star and eventually escape from its surface. In the presence of this process, the neutron star would lose mass gradually and emit a huge number of the dark photons into space. Due to the limitations of the present experimental techniques \cite{caputo2021dark,cline2024status}, so far 
no solid evidence has been found for the existence of the dark photons. If the mass of the dark photons lies within the directly detectable regions of future high-energy experiments or future astrophysical observations, theoretical calculations of the spectrum of the emitted dark photons may highlight the key open questions and directions for future research.

The dark photons created from the $nn \rightarrow VV$ process have a very small mass by assumption and can escape to infinity from the interior of the neutron star. During the escape process, the dark photons would lose kinetic energies to overcome the attractive gravitational potential, and thus a gravitational red-shift would occur \cite{fuller1996neutrino}. Following Refs. \cite{fuller1996neutrino,goldman2019bounds,goldman2025effects}, the red-shifted average total energy of the dark photon as measured by a distant observer can be evaluated by 
\begin{equation}
\begin{split}
E_{T}^{\infty} &\simeq E_{G}^{\infty} + E_{F}^{\infty}\\
               &\simeq e^{\Phi(r_{\text{em}})} E_{G} + e^{\Phi(r_{\text{em}})} E_{F}\\
               &\simeq e^{\Phi(r_{\text{em}})} m_n +  e^{\Phi(r_{\text{em}})}\frac{1}{2m_n}\Big(\frac{3\pi^2X_nN_a}{V} \Big)^{\frac{2}{3}},
\end{split}
\end{equation}
where $E_{F}^{\infty}$ ($E_{F}$) and $E_{G}^{\infty}$ ($E_{G}$) are the red-shifted (local) average Fermi energy and the total red-shifted (local) gravitational binding energy of the rest mass, respectively. The gravitational red-shift factor, which describes the change in energy due to the gravitational field, can be evaluated by $\sqrt{g_{00}(r_{\text{em}})} = e^{\Phi(r_{\text{em}})}$ \cite{shapiro2008black,glendenning2010special,glendenning2012compact,ferrari2020general,schutz2022first}. Here, $r_{\text{em}}$ is the radial coordinate of the emission point. Conventionally, the reference point for the gravitational potential energy is chosen as infinity [i.e. $e^{\Phi(r_{\infty})} \equiv 1$], where the gravitational field tends to be zero.

Figure \ref{fig07} shows the estimated dark-photon emission rates that are plotted in the mass-radius plane of the neutron star for different combinations of model parameters including the scalar boson mass $M_{\phi}$ and the coupling product $|g_1 g_2|$. As the gravitational mass of the neutron star increases, so does its density, creating a denser environment that accelerates the $nn \rightarrow VV$ process. As a result of this, more massive stars have higher dark-photon emission rates. The emission rate varies by several orders of magnitude across the parameter space displayed, ranging from approximately $10^{35}$ to $10^{40}$ dark photons per second. The variation is mainly due to both the assumed interaction strength and the internal structure of the neutron star, which is modeled using the MPA1 EoS. The emitted dark photons might be practically invisible through currently available experimental techniques, but could possibly lead to measurable signatures in future experiments, or even influence the long-term thermal evolution and mass (energy) loss mechanisms of the neutron star.

Figure \ref{fig08} shows the energy distribution of the emitted dark photons as viewed by a distant observer, taking into account both the local Fermi energy of neutrons and the gravitational red-shift caused by the compactness of the neutron star. Each energy value corresponds to a neutron star with a specific mass and radius, where the spectral pattern is generally determined by the gravitational potential at the emission location. As a result of the increased gravitational red-shift, more compact stars, which have higher mass-to-radius ratios, tend to produce dark photons with lower expected energies. Under the assumptions employed here (e.g., the mass of the dark photon $M_V \equiv 1$ keV), the peak energies of the emitted spectrum roughly fall within the range from 400 to 900 MeV. Although such dark photons are expected to interact with ordinary matter in an extremely weak manner, their spectral patterns may convey important signatures of both the neutron-star structure and the new physics parameters, offering a promising opportunity for future indirect detection efforts.

Figure \ref{fig09} shows the relationship between the average energy of the emitted dark photons, as measured from infinity, and the compactness parameter ($Cs \equiv R_s/R$) of the neutron star. The compactness parameter is defined as $Cs \equiv R_s/R$ \cite{haensel2007neutron}, where $R$ and $R_s$ are the physical radius and the Schwarzschild radius of the neutron star, respectively. As the compactness parameter increases, neutron stars become more compact, and then the gravitational red-shift becomes stronger, reducing the energy of emitted dark photons as measured from infinity. A more compact neutron star tends to have a higher average Fermi energy, but this is not a pronounced trend. Generally speaking, both the gravitational binding energy of the rest mass $E_{G}^{\infty}$ and the red-shifted total energy $E_{T}^{\infty}$ decrease with increasing the compactness value. Multiple internal configurations (i.e., different central densities and pressure distributions) could correspond to the same compactness value but would result in different characteristics of the energy spectrum and red-shift effects. The multivalued dependencies are easily recognized in the vicinity of the maximum mass configuration. This is because the compactness parameter is a global parameter, while the red-shifted energy is determined by both the local Fermi energy and the accumulation of red-shifts.

Figure \ref{fig10} shows the relationship between the red-shifted average energy of dark photons and the central density of the neutron star $\rho_c$. As mentioned above, the multi-valued dependencies suggest that the compactness parameter alone is not sufficient to fully determine the red-shifted energy spectrum of the emitted dark photons. A more comprehensive understanding requires further knowledge on the relationship between the energy spectrum and the central density $\rho_c$. The curve of the red-shifted total energy initially saw a sharp fall in the lower ranges of $\rho_c$ but then gradually leveled off, exhibiting a slow but steady growth trend in the upper ranges of $\rho_c$. Unlike $Cs$, each value of $\rho_c$ maps uniquely onto a single value on each curve of the red-shifted energy. Despite the theoretical clarity of figure \ref{fig10}, its limitation is that the central density $\rho_c$ is an internal quantity but not a directly observable quantity. In contrast, figure \ref{fig09} provides little information on the internal structure of the neutron star, but instead provides more observationally relevant information and thus is more closely related to experiments.

Some neutron stars in the early thermal evolution stages, such as Cassiopeia A, seemed to have experienced an anomalous cooling process \cite{heinke2010direct,shternin2011cooling}, which requires further explanations beyond the modified Urca process \cite{yakovlev2004neutron,page2006cooling,page2011rapid}. Previous studies proposed that dark sectors may contribute significantly to the anomalous cooling process \cite{hong2021cooling,shin2022dark,gardner2024probing}.
The $nn \rightarrow VV$ process introduces an additional channel for the mass (energy) loss of the neutron star in connection with dark sectors. These dark photons are assumed to couple minimally with ordinary matter and could travel freely in the interior of the neutron star, making them a highly efficient cooling agent, especially for young neutron stars. If the model parameters associated with the $nn \rightarrow VV$ process take appropriate values, the emitted dark photons could carry away thermal energies at a rate competitive with, or even greater than that of conventional channels. The expected surface temperature evolution of a certain neutron star would be modified if the $nn \rightarrow VV$ process was incorporated in its cooling analysis, potentially allowing constraints on the scalar mass $M_{\phi}$ and the coupling product $|g_1 g_2|$ through comparison with thermal observations of the neutron star.

\section{Summary and Outlook}

In this work, we examined a new $\mathcal{B}$-violating dineutron decay process, where two neutrons are allowed to convert into two dark photons ($nn \rightarrow VV$), in the dense neutron-dominated environment of neutron stars. Such a process may appear in some extensions of the SM, where additional hidden $U(1)_D$ gauge symmetries and new scalar bosons are introduced. Although the $nn \rightarrow VV$ process is significantly suppressed by the energy scale of new physics, it may still lead to measurable signatures in neutron stars due to their extremely neutron-rich environments. We extend the existing models slightly by incorporating effective interactions mediated by new scalar bosons to estimate the dineutron decay rate of the $nn \rightarrow VV$ process. The dineutron decay rate depends on numerous parameters, such as the product of the coupling constants ($|g_1 g_2|$), the mass of the dark photons ($M_V$), and the mass of the new scalar bosons ($M_{\phi}$). The values of certain parameters, such as the product of the coupling constants ($|g_1 g_2|$), and the mass of the dark photons ($M_V$), can be chosen to be consistent with collider experiments and precision spectroscopy constraints.

We adopted the MPA1 EoS to simulate the neutron star structure and used the RK4 approach to solve the TOV equations numerically. Based on them, we then evaluated the astrophysical consequences of the $nn \rightarrow VV$ process. We demonstrated that the gradual mass loss from the neutron star would occur over astrophysical timescales arising from the $nn \rightarrow VV$ process. Such a mass loss may cause small changes in the orbital period of binary pulsar systems. We set lower bounds on the mass of the new scalar bosons and the product of the coupling constants using precision timing data from a set of well-measured binary pulsar systems.

Besides the effects on the orbital period, we also estimated the possible energy spectrum of dark photons that escape from the neutron star due to the $nn \rightarrow VV$ process. Even though present detectors might be unable to detect such signals, future experiments or indirect astrophysical probes might provide a new avenue to explore these possibilities. The continuous mass (or energy) loss due to the $nn \rightarrow VV$ process may postpone the collapse of heavy neutron stars into black holes and may trigger the explosion of low-mass neutron stars. Furthermore, it may also accelerate the cooling processes of the neutron star and may influence the evolution phase of the gravitational wave signals. Due to the rich and varied signatures of the $nn \rightarrow VV$ process, a comprehensive analysis of these possibilities would lie beyond the scope of this work but would worth further exploration in future studies.

In summary, we consider massive or compact neutron stars to be highly powerful laboratories for the study of the $nn \rightarrow VV$ process. We show that currently available precision pulsar‑timing data could already probe the new physics energy scale well above a TeV. The constraints on the model parameters of new physics would directly be improved if the experimental upper bounds on the unexplained changes in the orbital period of binary systems were tightened.

This work also demonstrates how neutron stars can be used as powerful laboratories to explore new physics beyond the SM, and in particular the dark-sector physics. Despite being hypothetical, the $nn \rightarrow VV$ process is a prime example of how astronomical observations and particle physics models can be combined to place strong restrictions on the parameter space of new physics. These analyses could be further refined by future improvements in dark sector detection and pulsar timing observation.

\section*{Acknowledgement}
This work is supported by the National Natural Science Foundation of China (Grant No. 12104187), Jiangsu Provincial Double-Innovation Doctor Program (Grant No. JSSCBS20210940), and the Startup Funding of Jiangsu University (No. 4111710002). Y. H. thanks Dr. Yihao Yin and Dr. Leihua Liu for many useful conversations about General Relativity. \textbf{Disclosure Statement}: The authors declare that some portions of this manuscript share similar background motivation, methodological structure, and astrophysical assumptions with an earlier manuscript by the same first author (Y. H.) titled "\textit{Dineutron decay into sterile antineutrinos in neutron stars and its observable consequences}" in Ref. \cite{hao2023dineutron}. Although some portions of the text and derivations follow similar steps for continuity and completeness, the present work extends that analysis by focusing on a different decay channel involving dark photons, and includes new physical signatures, updated parameter constraints, and distinct particle phenomenology. All overlapping material and text have been included with the intent to guarantee that this manuscript is self-contained and scientifically coherent.

\bibliography{darkphoton}
\end{document}